# Using Big Data to Decode Private Sector Wage Growth

Dr. Ahu Yildirmaz
ADP
1 ADP Blvd.
Roseland, NJ, 07068
Ahu.Yildirmaz@ADP.com

## ABSTRACT

The U.S. labor market is dynamic and complex, and understanding wage data across different segments of the workforce is critical to providing policymakers and business leaders with actionable insights. There is no labor index that assesses the labor market performance at such a detailed level as the ADP Research Institute's Workforce Vitality Report (WVR). Drawing on the actual, aggregated and anonymous payroll data of 24 million Americans paid by ADP, the WVR looks at key dynamics and market indicators including wage growth, hours worked and turnover rate. Unlike other data sets, the WVR calculates wage growth of individual workers on a quarter-to-quarter basis, avoiding the deviations caused by various workplace occurrences, like when new workers are hired and older ones retire. In this paper, Dr. Ahu Yildirmaz, head of the ADP Research Institute, drills down into wage growth by industry, age, gender and income level, as well as for both job holders and job switchers. Using WVR data, Ahu walks through those factors contributing to overall shifts in wage growth, the future of the labor market and what this data means for today's U.S. workforce.

## 1. INTRODUCTION

With the constantly changing U.S. labor market and abundance of data in the market across different segments of the workforce, policymakers and business leaders look for a few specific indicators as measures of labor market strength. Each month, media, analysts and the business community look to a few reports, namely ADP's National Employment Report and the BLS jobs report, as indicators of economic health.

As it stands, most existing labor market indices are constructed at the national level focusing on a few metrics. The few indices that do provide more details measure either the overall performance for major metro areas or selected aspects of a local economy. Up until 2014, there was no comprehensive index that measured labor market dynamics at the industry, macro and regional levels, no existing indices to assess the labor market performance in intricate detail and no benchmark to measure human capital management for individual firms. So in 2014, ADP Research Institute launched the Workforce Vitality Report (WVR) to provide a deeper look at the labor market and wage situation.

Industrial, geographic and demographic characteristics of a certain labor market segment can look quite different from the national trend, so we would previously have to compile several different reports and data sets to understand the driving forces of labor market dynamics a. With the WVR, job seekers and employers are able to see how finely defined categories compare to various national and regional averages. It also enables organizations to adjust their policies in response to the changing market conditions.

Among the labor market indicators, the most followed data relates to wage and employment. The WVR tracks these economic barometers in one report to show the overall vitality of the labor markets.

## 2. METHODOLOGY

ADP provides payroll services for 24 million American private sector workers. This anonymous payroll data gives us insight into the workforce dynamics of one in six employees. We have two unique advantages in using this data. First, the data enables us to track the same firm and employee over time. Therefore, we distinguish different types of workers in the labor market: those who stay with the same firm (job holders) and those who change jobs (job switchers). Such a distinction enables us to measure how much a person or group grows their wages by switching, done so by computing the job switch rate and comparing pre and post job switch wages. Second, the ADP data provide employee and firm demographic variables, such as a firm's industry and size and an employee's age and gender. Human capital management (HCM) firms particularly find this useful because it gives them data vital to measuring their employees against various groups in the changing workplace.

One of the main differentiators is that ADP is able to track the wages of the same individual across time. Wage indicators available from other sources, including the BLS, calculate average wage growth based on a dynamic group of employees at the two ends of the time frame under consideration. ADP can follow the wages of a particular employee through time, rendering a much truer measurement of wage growth without dilution from constant flow of labor in and out of the market.

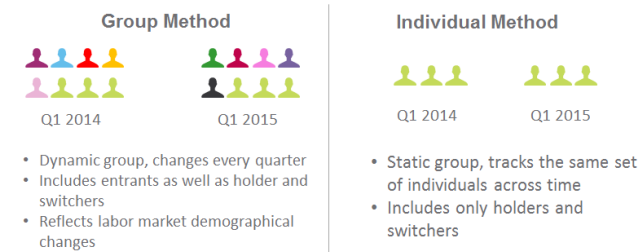

It tracks four other indicators in-depth:

**Turnover or job Switching rate:** The percentage of workers who successfully changed their jobs in consecutive quarters. In contrast to the separation and quit rates reported by the BLS, the turnover rate gives a clear indication of labor market conditions by isolating those workers who successfully changed their jobs and excluding those whose departure was due to retirement, company closure or who dropped out of the labor force. Because the majority of those who land a new job left the previous job voluntarily, the job switch rate increases when labor market conditions improve. However, the job turnover rate is not uniform across different industries. Rather, this component is of particular interest in distinguishing labor market conditions for different dimensions of workers to ascertain whether, for example, higher or lower wage workers are switching jobs, whether workers of a certain age or tenure on the job are switching or staying put, or whether workers in certain industries or certain regions of the country are switching jobs.

**Change in quarterly hours worked for job holders**: This concept characterizes the degree of labor utilization. When a specific labor market is sluggish, weak demand causes under-utilization of labor. When the specific labor market picks up, working hours increase as production expands. In addition, firms that employ high-skilled workers tended to retain their employees by cutting working hours during the downturn so as to avoid the search and rehiring costs when the economy rebounded. Hours are also cut as the demand for a firm's products weakens. As such, fluctuations in hours worked vary across industries with different concentrations of skilled workers.

**Change in average hourly wage rates of job holders:** In an improving labor market this change reflects employers' use of monetary incentives to retain talent and raise productivity. Changes in the turnover rate and hourly wage rates for job switchers are likely to lead to changes in the hourly wage rates for job holders. However, workers in a specific industry or at particular wage levels may be better positioned to bargain for higher wages sooner in the business cycle than workers in others.

**Change in average hourly wage rates of job switchers**: Change in hourly wage rate between the new job and the old job is a unique indicator not available from government sources. The initial wage offer is more sensitive to labor market conditions compared with wages of job holders. A higher wage offer is often a result of a tighter niche labor market where employers compete for talent. Like the job turnover rate, the ability of job switchers to grow their wages varies by age, tenure, gender, pay scale, industry and region. Seeing which group is able to command higher wages is of particular importance for HCM companies. For example, the wage change for minimum wage and high income switchers can be very different. While the former may be hard pressed to find better paying new jobs, the latter can more readily boost their wages by moving into new positions and trading up.

## DIMENSIONS CAPTURED BY THE ADP WORKFORCE VITALITY REPORT

1. Region: Northeast, Midwest, West and South

2. State: New York, New Jersey, Pennsylvania, Texas, Florida, California, Illinois, Washington, Michigan and Ohio

3. Industry (NAICS code): natural resources and mining (21), construction (23), manufacturing (31, 32, 33), trade and transportation (42, 44, 45, 48, 49, 22), information (51), finance/real estate/ (52, 53,), professional and business services (54, 55, 56), education & healthcare (61, 62), leisure & hospitality (71, 72), and other services except public services (81)

4. Firm size: 1-49, 50-499, 500-999, and 1,000 and above

5. Age: 16-24, 25-34, 35-54, and 55 and above

6. Gender: male and female

7. Full and part-time: full-time workers are defined as those whose weekly hours is greater and equal to 35

8. Wage tier: based on annual wage; less than 20K, 20K-50K, 50K-75K, and 75K and above (upper bound excluded)

9. Tenure: less than 3 years, 3-5 years, 5-10 years, and 10 years and above (upper bound excluded)

## 3. SELECTED FINDINGS

### WAGE OVERVIEW

By measuring individual workers who have remained in the same job over the past year, we are able to get a comprehensive glimpse into the wage growth situation. Unlike the BLS measure, ADP data allows us to calculate wage growth of individual workers and control for events that tend to skew labor data, such as when new workers are hired, older ones retire, and the mix of workers changes between full-time and part-time. If highly paid workers are leaving an industry to retire, for example, the aggregate wage growth will weaken because these are likely higher paid workers. An establishment could be replacing retiring workers with younger, less experienced workers who are earning less than the retiring workers. Alternatively, to keep costs down companies may opt not to replace the retiring high-priced workers at all. These changes could have the effect of constraining the average wage growth.

Tracking the same set of full-time workers leads to a truer picture of wage growth among those who are consistently employed. This set of full-time workers falls into two categories: job holders, those who stay in the same job, and job switchers, those who change jobs.

### ANNUAL WAGE GROWTH Q1 2016

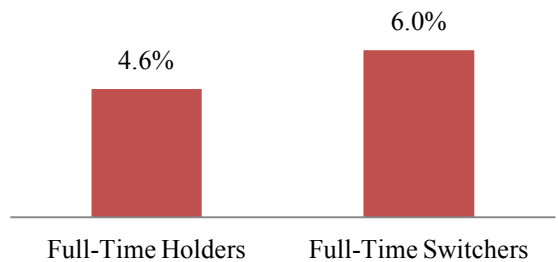

The richness of ADP data allows us to identify the segments who fared best in the workforce. For example, in Q1 2016, the data enables us to summarize wage trends across dimensions. The best wage growth is in the West, in information industries, among women, younger workers, those with little job tenure and those employed at the largest companies.

**Who Fared the Best at the Start of the Year Wage Growth from a Year Ago, Q1 2016**

| Age | 24 and younger | 9.5 |
|---|---|---|
| Industry | Information | 7.6 |
| Tenure | 2 years or less | 7.4 |
| Gender | Female | 4.0 |
| Size | 1,000 and | 5.2 |
| Region | West | 5.2 |

### TURNOVER RATE

ADP's definition of turnover differs somewhat from that of BLS. BLS turnover is calculated as (all separations)/(average employment) for the period under consideration. The numerator in BLS "all separations" includes quitting the labor market altogether and all job switching from one organization to another. ADP's definition is (number of jobs switched from one ADP client to another)/(All job holders, job switchers, new entrants).

Turnover rate has continued to increase steadily to 25.0 percent in Q1 2016. Considering the reasons for job switching, to trade up and because the job market seems accommodating, turnover is a good thing for the labor market. It is also beneficial for employers since they are looking to match jobs with the right talent.

Not surprisingly, turnover decreases as workers progress in their careers. Workers 24 and under had a turnover rate of 50.6 percent while workers 55 and over had a rate of 10.1 percent. Manufacturing has an average age of 43 according to BLS, so it is not surprising that the higher age and emphasis on highly skilled workers results in the lowest turnover rate of any industry, 12.7 percent. Leisure and Hospitality has the highest turnover rate, 45.4 percent.

The more skilled the industry, the less the likelihood that there will be turnover because workers and employees invest too much in training to hop jobs frequently. High turnover is not good for employers, but sometimes, it is a part of the process for employers to optimize their talent pool. Ideally, from employer's perspective, they would like to reduce turnover and increase the holders' wages in order to strengthen retention.

### AGE

The youngest job holders, under 25 years of age, saw their wages rise by 9.2 percent in the four quarters ending with the first quarter 2016. This is not surprising, since younger employees with less experience would command a much lower base wage, and any increase would be amplified due to a lower base. Job switchers in this demographic saw their wages rise by 11.1 percent. Most of the overall wage growth of 6 percent for job switchers can be attributed to the below 35 age group. Older workers are expected to be more experienced and hold more senior positions, so the incentive to change jobs is much smaller in this group. Even if they do change jobs, because of their higher base wages, the percentage increase is not that substantial as lower wage, younger workers.

## TRANSITIONING FROM PART-TIME TO FULL-TIME

The status of the job a worker switches from matters. If the change is from a full-time job to another full-time job, across all industries except Natural Resources and Mining, switchers experience moderate to high growth in their hourly wages. On the other hand, moving from a part-time to a full-time job across the industries generally results in a decrease in hourly wage. This may reflect that, when switching from part time to full time, the availability of benefits, more hours, increased take home pay and the stability of a more permanent count for a lot and may outweigh an hourly wage increase alone.

## WAGE GROWTH FOR JOB-SWITCHERS

| Industry | Full Time to Full Time | Part Time to Full Time |
|---|---|---|
| **ALL** | **6.0%** | **-3.1%** |
| Manufacturing | 3.2% | 1.9% |
| Construction | 3.8% | 3.8% |
| Natural Resources and Mining | -7.0% | -13.9% |
| Finance and Real Estate | 6.7% | -6.3% |
| Information | 7.5% | -9.3% |
| Professional and Business Services | 6.4% | -5.4% |
| Leisure and Hospitality | 10.1% | 8.9% |
| Education and Health Services | 5.1% | 0.2% |
| Trade, Transportation, and Utilities | 9.8% | 5.1% |

## 4. CONCLUSION

Wage growth has been the missing ingredient for full economic recovery, and in Q1 2016 it made a big improvement. Overall employment is rising, and wages are starting to pick up some momentum. More broadly, this is something we are starting to hear more about in the media. Wage growth is an indicator that the Federal Reserve regularly considers in its discussions and is an indicator relevant to everyday Americans.

However, until 2014, there was not yet a comprehensive, single report that aggregated all of this data. So, having access to the actual wage data, ADP Research Institute compiled its findings and has released the Workforce Vitality Report quarterly, free of charge. This data gives policymakers, economists and business people the opportunity to see an objective snapshot of the labor market and how it relates to their area of expertise.

Effective reporting is nothing new for us. ADP just celebrated the tenth anniversary of its National Employment Report and we have had deep roots in big data since before it was called "big data."

There is no other organization in the country with the data necessary to deliver insights into the health of the workforce with the depth and accuracy that ADP has. We also understand that there is a responsibility to use this data for good, and that is what we look to do by distributing our reports so that the public can have a fair, objective view into how the labor market is performing.

# 5. REFERENCES

ADP Workforce Vitality Index, March 2016
http://workforcereport.adp.com/